\documentclass[prd,tightenlines,twocolumn]{revtex4}
\usepackage{amsmath}
\usepackage{amssymb}
\usepackage{graphicx}
\usepackage{bm}
\usepackage{enumerate}
\usepackage{color}
\newcommand{\be}{\begin{equation}}
\newcommand{\ee}{\end{equation}}
\newcommand{\bea}{\begin{eqnarray}}
\newcommand{\eea}{\end{eqnarray}}
\newcommand{\ba}{\begin{eqnarray}}
\newcommand{\ea}{\end{eqnarray}}

\begin{document}

\title{Higher order string effects and the properties of the Pomeron}
\author{ 
 Dmitri Kharzeev$^{1,2}$ , Edward Shuryak$^1$ and Ismail Zahed$^1$}
 
\affiliation{$^1$ Department of Physics and Astronomy, \\ Stony Brook University,\\
Stony Brook, NY 11794, USA}
\affiliation{$^2$ Physics Department and RIKEN-BNL Research Center, \\
Brookhaven National Laboratory, \\
Upton, NY 11973, USA} 
%\date{\today}

\begin{abstract} 
We revisit the description of the Pomeron within the effective string theory of QCD.
Using a string duality relation, we show how the static potential maps onto the high-energy scattering amplitude that exhibits the Pomeron behavior.
Besides the Pomeron intercept and slope, new additional terms stemming from the higher order 
string corrections are shown to affect both the 
growth of the nucleon's size at high energies and its profile in impact parameter space. The stringy description also 
allows for an odderon that only disappears  in critical dimension. Some of
the Pomeron's features that emerge within the effective string description can be studied at the  future EIC collider. 

 \end{abstract}

\maketitle

%%%%%%%%%%%%%%%%%%%%%%%%%%%%%%%%%%%%%%%%
\section{Introduction}

Hadron-hadron scattering at high energy is dominated by the exchange of weakly interacting Pomerons and Reggeons with vacuum and meson 
quantum numbers respectively. Many decades ago, a 
description of this process within the effective Reggeon field theory was pioneered by Gribov and collaborators~\cite{Gribov:1973jg}.
%{Donnachie:1992ny}. 

A first principle approach to Reggeon physics in the context of weakly coupled
QCD was succesfully obtained by re-summing rapidity ordered Feynman graphs \cite{Kuraev:1977fs}.
The BFKL Pomeron~\cite{Kuraev:1977fs}   emerges in leading order through the re-summed two-gluon
ring diagrams with vacuum quantum numbers. Few non-perturbative approaches
to Reggeon physics in the context of QCD have also been attempted, ranging from
the stochastic vacuum~\cite{PIRNER} to a Reggeized graviton in holography~\cite{Brower:2006ea}.  
Here, we will be  interested in  the Pomeron as a semi-classical  stringy
instanton or in short the BKYZ Pomeron~\cite{Basar:2012jb}.

The BKYZ Pomeron carries intrinsic temperature  and entropy~\cite{Shuryak:2013sra},
i.e. $T=\chi/(2\pi b) $ with $\chi={\rm ln}(s/s_0)$ being the relative rapidity of the pair of  hadrons scattering at the
impact parameter $b$. As a result, a new phenomenon in hadron-hadron scattering at large
energy may take place when the intrinsic temperature of the string becomes comparable to
the Hagedorn temperature. A highly excited and long string with large energy and entropy 
becomes a string ball, which, when cut, can lead to the large multiplicity states
observed in hadron-hadron energies at collider energies.

In this paper we will focus on the systematics of the QCD effective string theory (EST),
when the hadron-hadron scattering at large $\sqrt{s}$ is dominated by exchange of weakly 
fluctuating  strings in a tube configuration. 
 In section \ref{sec_pot} we review what is known about QCD strings from two sources:
 (i) the  effective  string theory (EST) and (ii) numerical studies of lattice gauge theory.
Each is focused on the static potential between point-like  and heavy color charges.
In section \ref{sec_duality} we  use the string duality relation, and translate those results from the 
static potential to the Pomeron scattering amplitude. The phase of the Pomeron is addressed both
empirically and theoretically in the context of the BKYZ Pomeron in section IV. The evidence of the 
odderon from the time-like or $t>0$ region is reviewed in section V using the latest lattice results.
A charge odd analogue to the charge even Pomeron is identified as the stringy odderon.  In section
VI  and VII we discuss the empirical evidence for the size and shape of the Pomeron at collider energies,
and suggest their theoretical descrptions using the EST of the Pomeron. Our summay and 
conclusions are in section VIII. In the Appendix we briefly detail the unitarization of the string
amplitudes for fixed signature  in the EST.

%The main argument of this paper is that
%while the phenomenology of the static potential is only available indirectly, via Reggeons and heavy quark spectra, the %profile of the Pomeron $K(b)$ can be obtained directly, by 
%Fourier-Bessel transform of the experimental elastic amplitude. Furthermore,
%this profile contains the relevant potential in the exponent.  Therefore,
%the proposed approach can be potentially much
%more accurate, hopefully allowing to relate the LHC elastic scattering data 
%t%o effective theory of the QCD strings.

%The phenomenological discussion is split into two section. In \ref{sec_size} we
%discuss the proton size and cross section, and in section  \ref{sec_shape}  the details of its shape.

\section{Effective string theory and the static potential } \label{sec_pot}
 
At large distances, the leading contribution to the static and heavy quark-antiquark
potential $V_0(r)$ in pure Yang-Mills theory is the famous linear potential

%The leading order confining potential $V_0(r)$ at large distances $r\rightarrow \infty$
%in pure gauge theory is just the famous linear potential

\be 
\label{0}
V_0(r) =\sigma_T r 
\ee 
%with the coefficient known as the string tension. 
with $\sigma_T$ as the fundamental string tension. 
In QCD with light quarks
this behavior is only valid till some distance due to screening by light quarks
in the form of two heavy-light mesons. This effect will be ignored below.
%due to crossing with the two heavy-light mesons
%state: we will ignore light quark phenomena below.

Long strings are described uniquely by the Polyakov-Luscher action~\cite{Polyakov:1986cs,Luscher}
(an expanded form of the Nambu-Goto action) in the form
%
% the Nambu-Goto action
\be 
\label{1}
S=-\sigma_T \int_M d^2x \,(1+\partial_\alpha X^i \partial^\alpha X_i) 
\ee
The integration is over the world-volume  of the string $M$ with embedded coordinates $X^i$
in D-dimensions. The first contribution is the area of the world-sheet, and the second contribution
captures the fluctuations of the world-sheet in leading order in the derivatives.

Since the QCD string is extended and therefore not fundamental, its description in terms
of an action is  ``effective" in the generic sense, organized in increasing derivative contributions
each with new coefficients.  These contributions are generically split into bulk $M$ and boundary 
$\partial M$ terms. The former add pairs of derivatives to the Polyakov-Luscher action. The first of such 
such a contribution in the gauge fixed as in (\ref{1}), was proposed by Polyakov~\cite{Polyakov:1986cs}

\be 
\label{2}
+\frac 1\kappa \int d^2x\, \left(\dot{\dot {X}}^\mu\dot{\dot{X}}_\mu +2\dot{X}^{\prime \mu}\dot{X}^\prime_\mu+
X^{\prime\prime\mu}X^{\prime\prime}_\mu\right)
\ee
which is seen to be conformal with %$\kappa$ 
the dimensionless extrinsic curvature.
Higher derivative contributions are restricted by
Lorentz (rotational in Euclidean time) symmetry. The boundary contributions are also restricted 
by symmetry. The leading contribution is a constant $\mu$, plus higher derivatives. We will only
consider the so-called $b_2$ contribution with specifically

 %The boundary action starts with the
%boundary constant and we will only discuss in particular the so called $b_2$ term

\be 
\label{3}
S_b=\int_{\partial M} d^2x \left (\mu +b_2\, (\partial_0\partial_1 X^i)^2\right)
\ee

All the terms in  (\ref{1}-\ref{3})  contribute to the static potential (\ref{0}). The first contribution stems from the string
vibrations as described in the quadratic term (\ref{1}),

\be
\label{4}
\sigma_Tr\left(1+\frac {V_0}{\sigma_Tr^2}\right)
\ee
It is Luscher universal term with $V_0=-\pi/12$ in 4-dimensions~\cite{Luscher}.  Using string dualities,
 Luscher and Weisz~\cite{Luscher:2004ib} have shown that the next and higher contribution is universal

\be
\label{5}
\sigma_Tr\left(1+\frac {V_0}{\sigma_Tr^2}-\frac 18 \left(\frac{V_0}{\sigma_Tr^2}\right)^2\right)
\ee
These contributions are part of a string of contributions re-summed by~Arvis \cite{Arvis:1983fp} 
using the Nambu-Goto action

 \be    
\label{6}
V_{\rm Arvis }(r)=\sigma_T r \left(1-{\pi \over 6} {1 \over \sigma_T r^2}\right)^{\frac 12}
 \ee
For further discussion of the static $Q\bar Q$ potential stemming from the EST we refer to recent work in~\cite{Aharony:2010db}.  To order $1/r^4$ all the bare contributions to the potential are known

\be
\label{VB1}
V(r)\approx \sigma_T r -\mu-\frac {\pi D_\perp }{24 r}-\frac{\pi^2}{2\sigma r^3}\left(\frac{D_\perp}{24}\right)^2 +{\tilde b_2 \over r^4} +...
\ee
with general number of transverse dimensions $D_\perp=D-2$.
The $\mu$ term receives both perturbative and non-perturbative  contributions.
The former are UV sensitive and in dimensional regularization renormalize to zero,
as we  assume throughout. The latter are not accounted for  in the conformal  Nambu-Goto string, 
but arise from the extrinsic curvature term (\ref{3}) in the form~\cite{Hidaka:2009xh,Qian:2014jna,footnote}.

\be
\label{VB2}
\frac{D_\perp}4\sqrt{\sigma\kappa}\rightarrow \mu
\ee
%\fbox{what is kappa?}
Note that this contribution amounts to a negative boundary mass term in 
(\ref{3}), and vanishes for $D=4$ spacetime-dimensions. It 
 is finite for $D_\perp> 2$ in the holographic AdS/QCD approach. The third and fourth contributions in (\ref{VB1}) 
are Luscher and Luscher-Weisz universal terms in arbitrary dimensions, both reproduced by expanding
Arvis potential; see \cite{Petrov:2014jya} for a related discussion of the role of Luscher terms in the Pomeron structure. The last contribution is induced by the derivative-dependent string boundary contribution (\ref{3}). 
(Note that even if the string ends are constant in
external space, they still may depend on the  2-dimensional  coordinates on the worldvolume of a vibrating string.)

 %The third and fourth contributions are Luscher universal terms, in which we keep arbitrary number of transverse %dimensions $D_\perp=D-2$, which is 2 in the ordinary
 %space, but 3 in the holographic AdS/QCD approach to the Pomeron, see below.
 %These two terms in (\ref{VB1}) are of course
%reproduced by the expansion of the Arvis potential. 
%The last term is induced by
%derivative-dependent string boundary term. 
%(Note that even if the string ends are constant in
%external space, they still may depend on 2-dimensional  coordinates on the worldvolume of a vibrating string.)
% Nambu-Gotto string with the exception of the zero point energy.  

We will not discuss the extensive holographic studies of the EST  and related  potential~\cite{Aharony:2010db}, 
but proceed to lattice simulations of the heavy-quark potential. These studies have now reached 
a high degree of precision, shedding  light on the relevance and limitation of the string 
description.  In a recent investigation by Brandt~\cite{Brandt:2017yzw} considerable accuracy was
obtained for the potential at zero temperature and for pure gauge SU(2) and SU(3) theories.
As can be seen from Fig.~3 in~\cite{Brandt:2017yzw}, the inter-quark potential is described to
an accuracy of one-per-mille, clearly showing that both  Luscher's universal terms, $1/r, 1/r^3$
are correctly reproduced  by the numerical simulations. Indeed, for  $r/r_0>1.5$ (or $r>0.75\,$  fm 
for Sommer's parameter $r_0=0.5\,$ fm) these two contributions  describe the potential
extremely well.

Expanding further to order  $1/r^5$, or keeping the complete square root in  Arvis potential
(\ref{6}), would not improve the agreement with the lattice potential, since the measured 
potential turns up and opposite to the expansion. Brandt  lattice simulations~\cite{Brandt:2017yzw}
have convincingly demonstrated that the next correction is  of order $1/r^4$ with the opposite sign. 
The extracted contribution fixes the $b_2$ coefficient   in (\ref{VB1}) as

\be 
\tilde b_2= - {\pi^3 D_\perp \over 60 } b_2 
 \ee
with the numerically fitted values

\bea
b_2^{SU(2)} \sigma_T^{3/2} = && -0.0257 (3)(38)(17)(3) \nonumber\\
b_2^{SU(3)} \sigma_T^{3/2}=&&-0.0187 (2)(13)(4) (2) 
\eea
(for the details and explanation regarding  the procedure and meaning of the errors we refer to~\cite{Brandt:2017yzw}).
Note that the overall contribution of this term to the potential is positive.
% two minuses which cancel each other, so the sign of this last term in the potential is positive.
 What this means  is that at $r\approx r_0=0.5\,$ fm the static potential 
contains a wiggle, visible however only with a good magnifying glass since its relative magnitude is  $10^{-3}$. We will
now explain how this impacts the scattering amplitude of color singlet dipoles at large $\sqrt{s}$.

\section{From the static potential to the  Pomeron} \label{sec_duality}

The (Euclidean) world-volume of the string for a static and heavy inter-quark  potential is a rectangle with the size
$\hbar/T$ in time and $r$ in space. It is assumed that $r \ll \hbar/T$ and the string is not excited.
In contrast, the  BKYZ Pomeron ~\cite{Basar:2012jb} is derived from a stringy instanton
with the world-volume of a shape of a ``tube", with a mean circumference $\beta=2\pi b/\chi$
and length $b$, the impact parameter. In this case a string is little excited when 
it is long in space, $b \gg \beta$, the opposite of the condition above.

As discussed in the Appendix of~\cite{Shuryak:2017phz}, one can 
map the two problems at hand via some duality relation, by exchanging time and space, and also by
adding another mirror image of a potential and match the boundary conditions. 
In this case the two partition functions of the string and its excitations become identical. 
The explicit transformation is 

\be 
2b \leftrightarrow {\hbar \over T}, \,\,\, \,\, \beta \leftrightarrow 2r
\ee 
Assuming the correspondence between the potential and the Pomeron is exact
% through the 
%exchange $2b\leftrightarrow \beta$ established in~, %
we can map the potential (\ref{VB1}) onto
the Pomeron scattering  amplitude in in impact parameter space  as

\be
\label{VB3}
{\cal A}{(\beta, b)}\approx 2is\,{\bf K}\approx 2is \,e^{-{\cal S}(\beta,b)} \ee
with explicitly

\bea
\label{SPOT}
{\cal S}=&&+\sigma\beta b-2\mu b-\frac{\pi D_\perp}{6}\frac b\beta\nonumber\\
&&-\frac{8\pi^2}{\sigma}\frac{b}{\beta^3}\left(\frac{D_\perp}{24}\right)^2 
-{ 2^5 b \tilde b_2\over \beta^4}
\eea
with $\sigma=\sigma_T/2$ and $2\pi\sigma_T=1/\alpha^\prime_{\mathbb R}$. 
Now, we can recall the parameters of the ``tube" and set
 $\beta=2\pi b/\chi$, with the shorthand notation 
  
\be 
\chi={\rm ln}(s/s_0)
 \ee

 Following this substitution, one observes that the leading and subleading  terms
have very different roles and  energy dependence. For instance, the leading two contributions 
with a rapidity dependence in (\ref{VB3}) are

 \be
 \label{VB4}
e^{\chi\frac{D_\perp}{12}-\frac{b^2}{4\chi\alpha_P^\prime}}
\ee
with the  Pomeron intercept of $\Delta=\frac{D_\perp}{12}$,
and the term responsible for the Pomeron slope. The latter yields the famous Gribov diffusion, 
at the origin of the large-distance growth of the hadrons according to the
``diffusive law"
\be 
b^2 \sim \chi ={\rm ln}\left({s\over s_0}\right)
\label{eqn_diff}
 \ee  
which exists equally for perturbative gluons and strings.

Furthermore, one should recognize that the stringy Pomeron approach exists in two versions, the flat space  and the holographic
ones. In the former case the space has two flat transverse directions $D_\perp=2$,
while in the latter the string  also propagates in the third  and curved dimension. Since Gribov diffusion
also takes place along this coordinate, identified with the ``scale" of the incoming dipoles,
the expressions we will use are a bit modified from the standard expressions.
One such effect, derived for the  the  BKYZ Pomeron in~\cite{STOFFERS}, is the modification of the Pomeron intercept 
\be 
%\label{NEWD}
{D_\perp \over 12} \rightarrow {D_\perp \over 12}\left(1-{3(D_\perp -1)^2 \over 2D_\perp \sqrt{\lambda} }\right) 
\label{eqn_Delta_P}
\ee
 Here $D_\perp=3$ and $\lambda=g^2N_c$ is the 't Hooft coupling, assumed to be  large. 
 In the range of $\lambda=20-40$,  (\ref{eqn_Delta_P})  is in the range 0.14-0.18. For the numerical analyses to
 follow, we will use for the Pomeron intercept the value $\alpha_{\mathbb P}(0)-1=\Delta_{\mathbb P}=0.18$.
 (This  happens to be  not far from the flat space value of $\frac 16=0.166$.) 

At the end of this section, let us discuss the way we fix the absolute units used in this work.
There is a dilemma,  well known in the literature.  One textbook approach is to rely 
on the slope of the mesonic/baryonic trajectories $\alpha^\prime_{\mathbb R}=1/(2 \pi \sigma_T)\approx 0.9 \, {\rm GeV}^{-2}$,
related with the fundamental tension of the open strings. If so, the closed strings glueballs
should have trajectories with $\alpha^\prime_{\rm glueballs}=\alpha^\prime_{\mathbb R}/2\approx 0.45 \, {\rm GeV}^{-2}$
as the tension is doubled. Using those canonical values is one option to fix the scale.
However, multiple fits to the Pomeron produce different and much smaller values for $\alphaÕ_{\mathbb P}$. The earlier fits
 in~\cite{Burq:1982ja} yield $\alpha^\prime_{\mathbb P}=0.14\pm 0.03\,{\rm GeV}^{-2}$, while the later ones  
 in~\cite{Donnachie:1992ny} yield $\alpha^\prime_{\mathbb P}=0.25\,{\rm GeV}^{-2}$. We note that the string tension
is warped in holography, and one would expect an effective $\sigma_T$ and therefore an effective $\alpha_R^\prime$ on average. This issue will not be addressed here.

\begin{figure}[h!]
\begin{center}
\includegraphics[width=7cm]{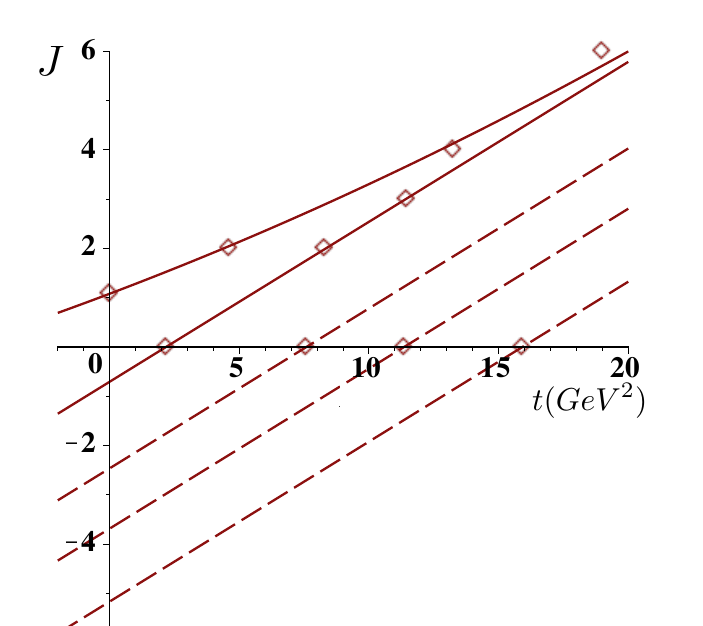}
\caption{Positive C-parity glueball according to lattice gauge
simulations  \cite{Meyer:2004gx}. The lines are our fit to Regge trajectories
\cite{Shuryak:2013sra}. }
\label{fig_Cpositive}
\end{center}
\end{figure}

In our previous paper \cite{Shuryak:2013sra} we used the results of lattice gauge
simulations ~\cite{Meyer:2004gx}, as reproduced in Fig.\ref{fig_Cpositive}, which
explains this deviation by the fact that the leading trajectory -- apparently unlike 
the second one -- is not linear but has a quadratic term. Our fitted value for $\alphaÕ_{\mathbb P}=0.20\, {\rm GeV}^{-2}$.
We will be using this value to fix the units in this work. %Using the observed Pomeron slope

%or the observed slope of the 

%The effects induced by the higher  order contributions  will be discussed  below.

  \begin{figure}[htbp]
\begin{center}
\includegraphics[width=7cm]{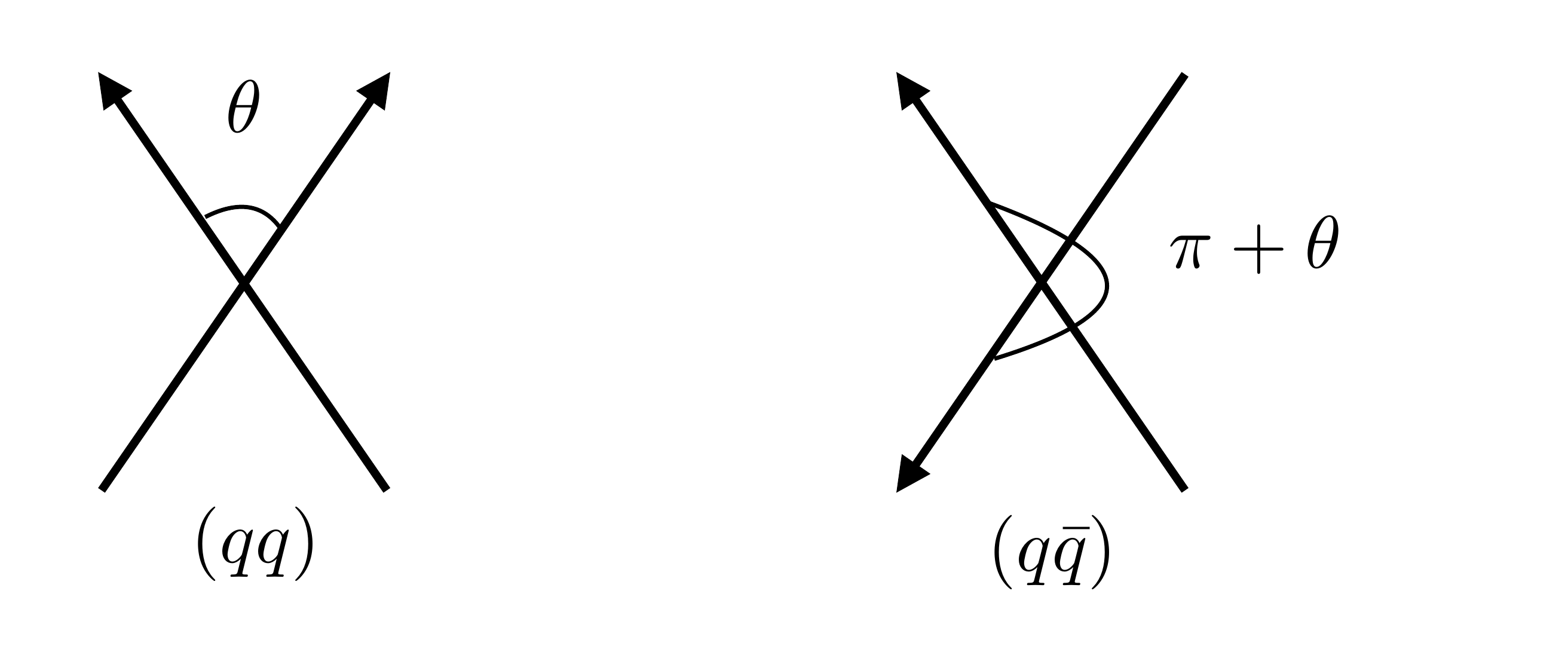}
\caption{Euclidean rendering of the $qq$ scattering amplitude at relative angle $\theta$ and the $q\bar q$ 
scattering amplitude at angle $\theta+\pi$. The continuation to Minkowski space follows through $\theta\rightarrow -i\chi$
with the latter as the charge conjugate of the former. The relevance of this construction for the Pomeron and the Odderon is detailed in the text.}
\label{fig_crossed}
\end{center}
\end{figure}

 %E new
   \section{The phase of the Pomeron  }
    
Experimentally, the Pomeron scattering amplitude exibits both a real and imaginary part. 
The real part is measured at two locations:  at  small $t\approx 0$, by observing the
interference with the weak Coulomb scattering, and   at the location  of 
the diffractive peak where the imaginary part vanishes.  For the interference measurement,
the results are expressed in terms of the so called $\rho$ parameter 
\be
 \rho={{\rm Re}({\cal A}(s,t=0)) \over {\rm Im}({\cal A}(s,t=0)) }
\ee
  The recent TOTEM data~\cite{Antchev:2016vpy} give
  \be 
\rho(\sqrt{s}=8\, {\rm TeV})=0.12\pm 0.03 
\ee
%The simplest  combination of the data is captured in the   so called ``standard fit"
%for the scattering amplitude
%  \bea
%&& {\rm arg}({\cal A})={\pi \over 2} - {\rm arctan}(\rho)\nonumber\\
%&&+ {\rm arctan}\left({|t|-|t_0| \over \tau}\right) -{\rm arctan}\left({-|t_0| \over \tau}\right) \nonumber\\
%\eea
%  with the parameters $t_0 = -0.50\,$ GeV$^2$ and $\tau = 0.1\,$ GeV$^2$. 
The textbook 
description of the Regge  scattering amplitudes relates the $\rho-$parameter with the signature factor 
%in the Pomeron
%channel is of the  form  (see for instance the review~\cite{Donnachie:1992ny}), 
%   
%
% \be 
%\label{DL}
%{\cal A }_{\mathbb P}= 2is[3 \beta_{PNN} F(t)]^2 (-i s \alpha')^{\alpha(t)-1} 
%\ee
which for small $t$ is captured by the phase factor $e^{-\frac i2\pi \Delta_{\mathbb P}}$. It is small
if $ \Delta_{\mathbb P}$ is small, in agreement with data.

We now note how similar terms appear in the stringy amplitudes, starting from the positive definite charge 
signature $C=+1$, the Pomeron. Such a  phase appears when analytically continuing the Euclidean scattering amplitude
at angle $\theta$ to $\pi+\theta$ for the cross channel, as illustrated in Fig.~\ref{fig_crossed} for quark-quark  scattering.
The result after integrating over the impact parameter and keeping only the two leading
contributions in (\ref{SPOT}) is for the $C=+1$ signature

\bea
\label{TPLUS}
&&{\cal A}^+(t,s)\approx 2is\left(\frac \pi 2g_s a_D\right)^2\nonumber\\
&&\left(\left({\rm ln}(s)\right)^{1-\frac{D_\perp}2}\,s^{\alpha_{\mathbb P}(t)} +
\left({\rm ln}(-s)\right)^{1-\frac{D_\perp}2}\,(-s)^{\alpha_{\mathbb P}(t)}\right)\nonumber\\
\eea
with $\alpha_{\mathbb P}(t)=\Delta+{\alpha^\prime_{\mathbb P}} t$. At large $\sqrt{s}$ and small $t$, (\ref{TPLUS}) is
also characterized by an extra phase $e^{-\frac i2\pi\Delta}$ in agreement with textbooks. 
However, we note that while the pre-exponential factors with ${\rm ln}(s)$
disappear in 4-d flat space, i.e. $D_\perp=2$, as in the textbooks, they 
do not for  $D_\perp> 2$ as is the case of the  holographic model
(see (\ref{eqn_Delta_P})). The additional pre-exponential factors
stem in this case from the diffusion in the additional holographic direction,
and are thus completely necessary. Their inclusion through the branch ${\rm ln}(-s)\equiv {\rm ln}(s)+i\pi$ produces
a correction to the phase
which is asymptotically  subleading at infinite $\chi$, but at current energies is about -15\%  (it enters
with the opposite sign to the standard phase.) So, accurate measurements of the energy dependence of the phase can be sensitive to Gribov diffusion in the 5-th dimension.

  % The stringy Pomeron however has a different prediction due to prefactor
 %  \be \left({1 \over 1+i\pi/log(s)}\right)^{D_\perp/2} \approx 1-  i {\pi \over log(s)}{D_\perp \over 2}\ee
  % which generates a subleading at $s\rightarrow \infty$ but numerically 
 %  comparable phase. 

 \begin{figure}[h!]
\begin{center}
\includegraphics[width=6cm]{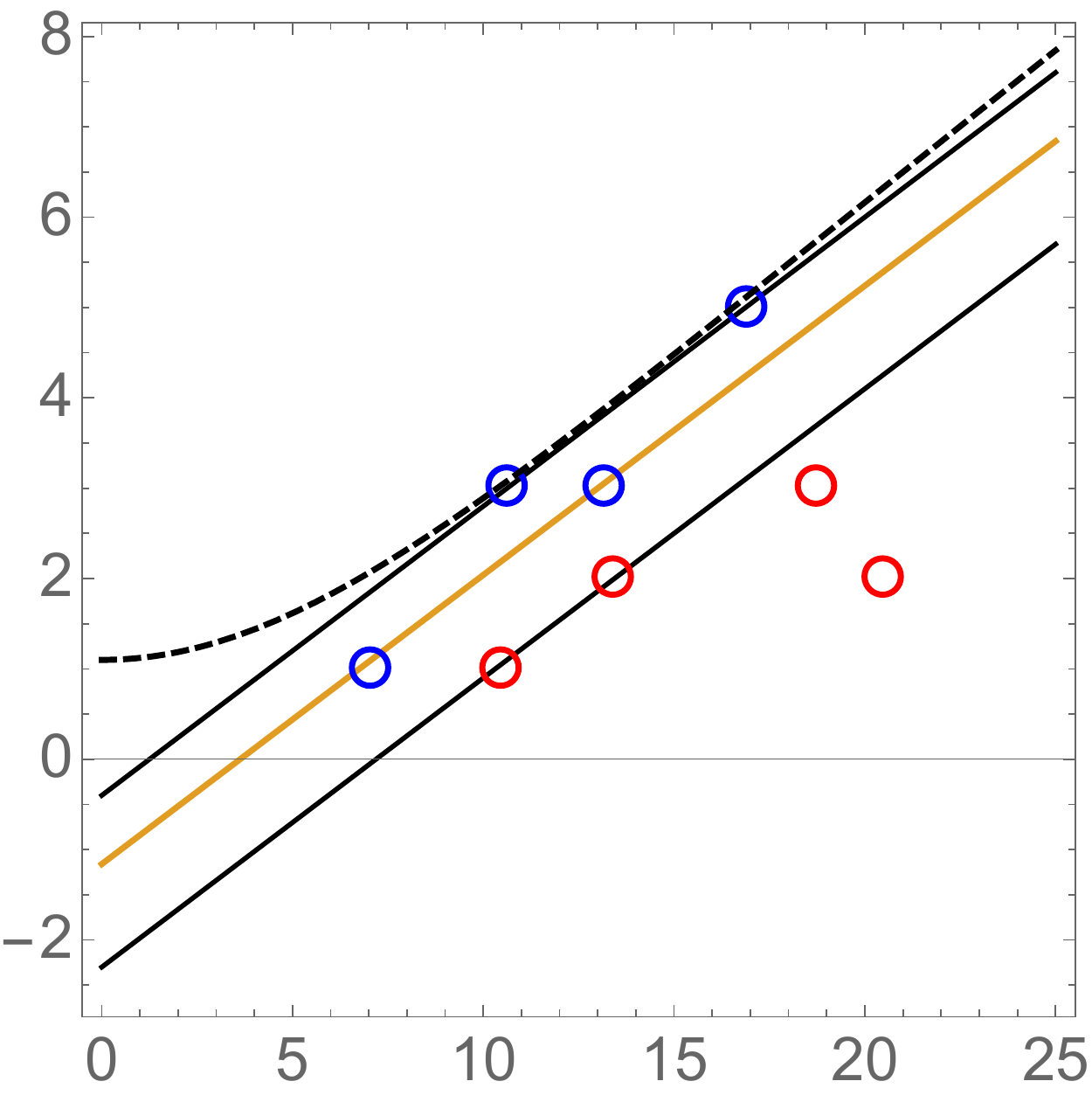}
\caption{$C=-1$ parity glueballs on a Regge plot, showing 
 their angular momentum $J$ versus their squared mass $M^2 ({\rm GeV})^2$.
  The two upper (blue) points and lines are for the negative spatial parity $P=-1$
  gluballs, and the lower (red) ones are for the $P=+1$.  The masses are from lattice gauge
simulations \cite{Meyer:2004gx}.
The lines shows the hypothetical behavior of the Regge trajectories.}
\label{fig_odd}
\end{center}
\end{figure}

  \section{ The Odderon} 
  
%Standard Regge literature requires the  odderon to be vector-like. It is expected to
%have the same amplitude, and in particular a similar phase factor to the Pomeron
%  
% \be 
%(-i s \alpha')^{\alpha_O(t)-1}  
%\ee
%where the trajectory $\alpha_O(t)$ is of course a different one.

Let us start with the phenomenology of glueballs. 
One can get some information about the odderon Regge trajectory
following the same procedure as we used e.g. in~\cite{Shuryak:2013sra}.
Specifically, this consists in:  1/ plotting the  known masses (squared)
of the appropriate glueballs; 2/    connecting  those by plausible lines, to learn the 
relevant slope of the trajectories; 3/  extrapolating the upper trajectory to $t=0$.
The corresponding plot for  the masses of  glueball with positive C parity, taken
from the lattice study by Meyer~\cite{Meyer:2004gx}, is shown in 
Fig.~\ref{fig_Cpositive} for completeness.

%Fig.~5 of~\cite{Shuryak:2013sra},
%and there is no need to reproduce it here. 

It is straightforward to extend it to glueballs with
 negative $C$-parity  as well, as we show in Fig.~\ref{fig_odd}.
  First, we note that there are 4 glueballs with negative $P=-1$ (vector-like) and 4 with positive
  $P=+1$ (axial-like) spatial parity. As seen from this figure, it appears that
  6 of those 8 states make three pairs defining straight and parallel lines with the same (within errors)
  slope, namely $\alpha^\prime=0.32\, {\rm GeV}^2$. Furthermore, we note that the lowest (the leftest)
  vector $J^{PC}=1^{--}$ state seems to belong to the second trajectory. This feature is the same as for the positive $C$ parity case, in which the lowest
 scalar $0^{++}$ also belongs to the second trajectory. We take both features as
 indications that the trajectories are correct as drawn here.

 Now, with only two states on the upmost trajectory, one can only define uniquely a straight line.
 Extrapolation of those to $t=0$ 
 would indicate that the odderon intercept $\alpha_{\mathbb O}(0)$ is negative. If so, its contribution 
 is  suppressed by at least one power of $s$ compared to the Pomeron, or 
 $\sqrt{s}$ compared to the  mesonic Regge trajectories: 
  in this case it is way too small to be observed at 
 RHIC/Tevatron/LHC energies. 
 
 However, we know from experiment
 (the scattering data at negative $t$ for the Pomeron) as well as from the
  positive $C$-parity plot (in which one knows 3 states, $J=2,4,6$) 
  that the upper Pomeron trajectory is not linear but curved. One may speculate
  that the same feature would hold  for the odderon as illustrated by 
the speculative dashed line in   Fig.~\ref{fig_odd}.

On the theoretical side, the stringy description of the Pomeron also allows for
a stringy odderon. The dipole-dipole scattering with negative signature follows 
the same reasoning as that for the positive signature in (\ref{TPLUS}), with now the result

\bea
\label{TMINUS}
&&{\cal A}^-(t,s)\approx 2is\left(\frac \pi 2g_s a_D\right)^2\nonumber\\
&&\left(\left({\rm ln}(s)\right)^{1-\frac{D_\perp}2}\,s^{\alpha_{\mathbb P}(t)} -
\left({\rm ln}(-s)\right)^{1-\frac{D_\perp}2}\,(-s)^{\alpha_{\mathbb P}(t)}\right)\nonumber\\
\eea
 At large $\sqrt{s}$ and small $t$, the amplitude (\ref{TMINUS}) is only parametrically
suppressed from (\ref{TPLUS}) by ${\rm tan}(\pi\Delta/2)$. Remarkably, for the critical 
bosonic string with $\Delta=D_\perp/12=2$, the odderon amplitude vanishes identically.
This is not the case, for the holographic string with $2< D_\perp\leq 3$, where the suppression
is parametrically small (especially for our empirical value of $\Delta_{\mathbb P}=0.18$), 
but with an odderon with the same intercept as the Pomeron.
The inclusion of the higher stringy corrections should not alter qualitatively these observations.

 Summarizing this section: the  negative $C$-parity glueballs hint towards the  existence of
 a set of Regge trajectories, with a slope different from any other set. If the
 leading trajectory is curved, the location of its intercept $\alpha_O(0)$ 
 is likely to lie between zero and one.  The odderon drops out from the critical
bosonic string, but otherwise carries a similar intercept as the Pomeron for the
non-critical and holographic string. While its relative contribution is  small, its relyable observation would also be an indication to diffusion in the 5-th dimension.

\begin{figure}
\begin{center}
\includegraphics[width=6cm]{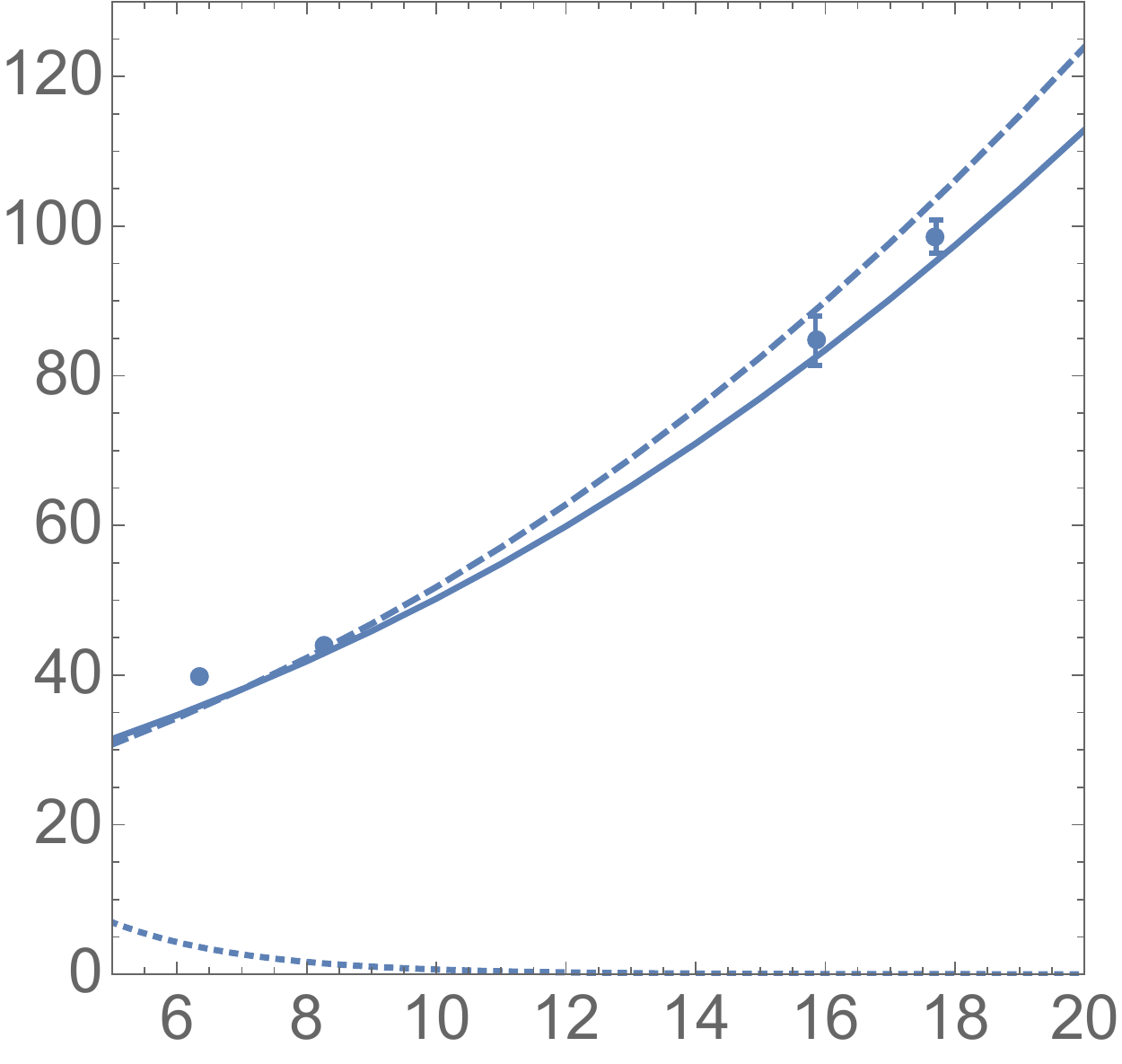}
\includegraphics[width=6cm]{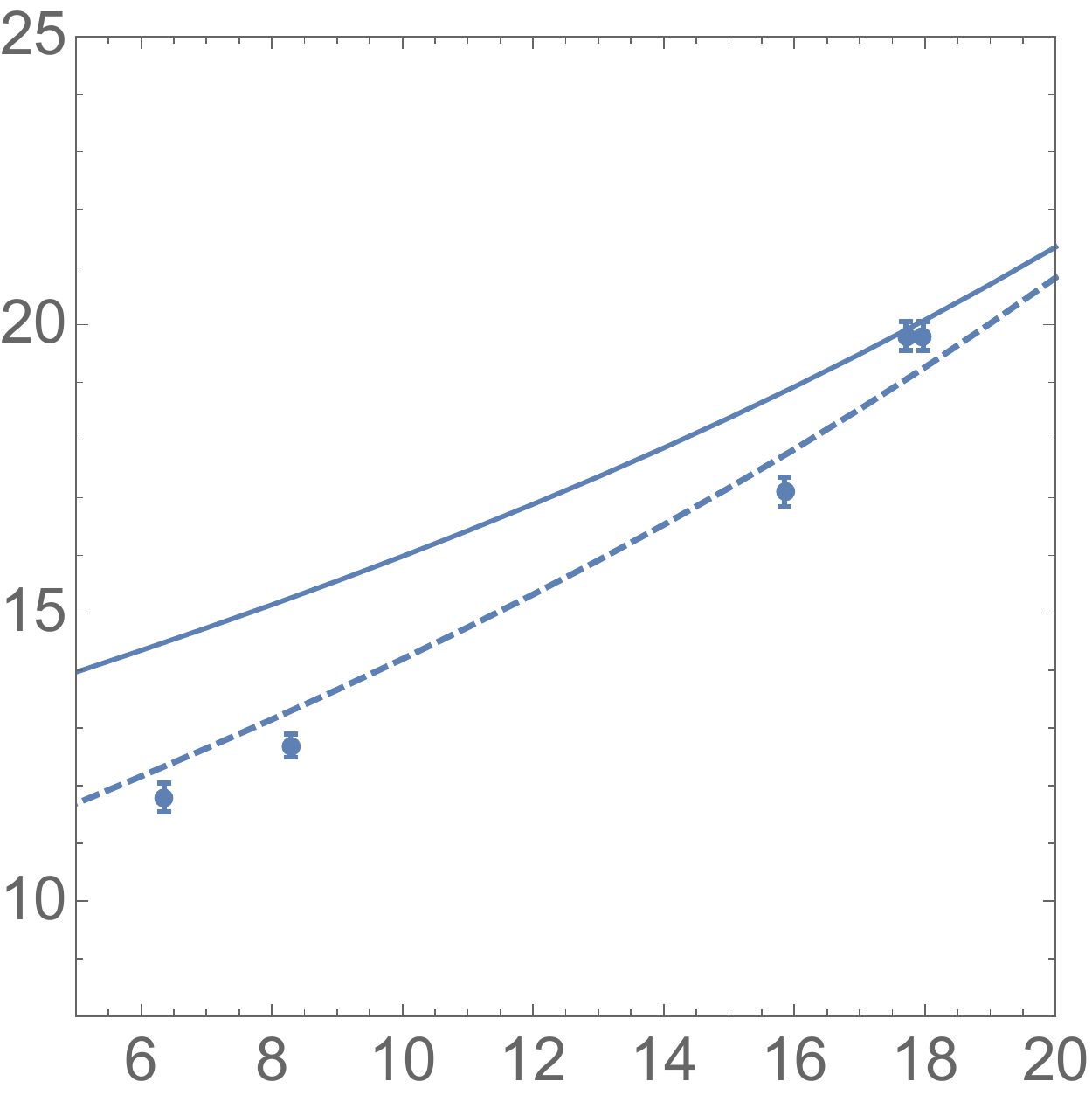}
\includegraphics[width=6cm]{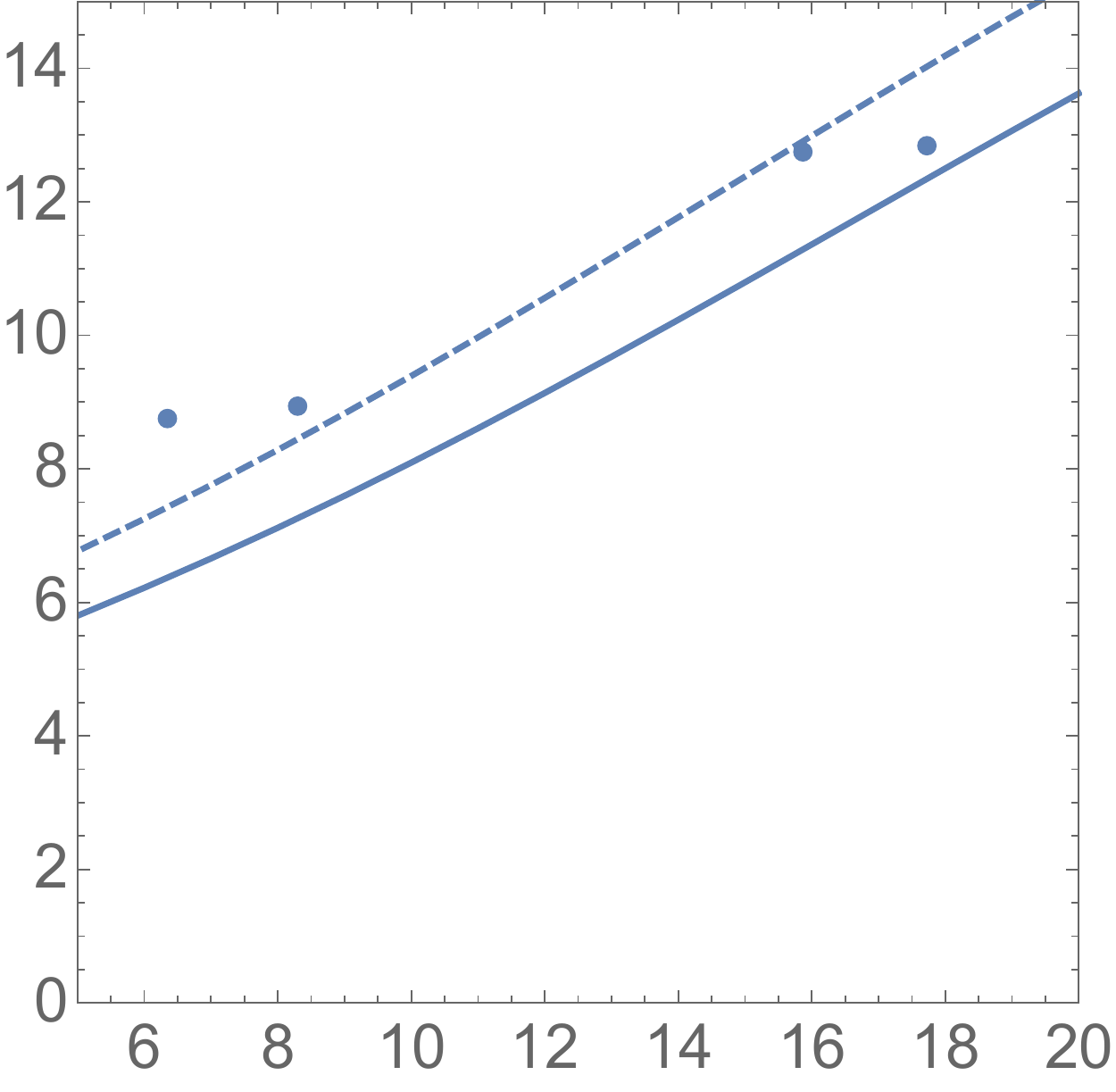}
\caption{The upper plot shows $\sigma_{\rm tot}$ in $mb$ versus the log of the collision energy $\chi={\rm ln}(s/s_0)$. 
The left-side (low energy) data points at $\sqrt{s}=27,63\, {\rm GeV}$  are from the old ISR measurements, and the three right-side points, for 
$\sqrt{s}=2.76, 7, 8 \, {\rm TeV}$ are from the TOTEM measurements. The dotted line in the lower plot indicates the contribution of the Reggeons other than the Pomeron (from the PDG fit.)
The middle plot shows the elastic slope  $B$ (${\rm GeV}^{-2}$), and the lower plot 
is their ratio. The curves are for the profile shown in Fig.~\ref{fig_profile_from_profile6}.
%solid curves at all plots are for the parameters $A=7.\,{\rm GeV}^{-2}$, $\mu =-0.04\, {\rm GeV}$, and the dashed ones are for 
% $A=4.5\,{\rm GeV}^{-2}$, $\mu =0$. 
}
\label{fig_pheno_from_profile6}
\end{center}
\end{figure}

\section{The Pomeron shape  }

\subsection{Phenomenology}

The main information we have about the shape of the scattering amplitude $\Omega(s,b)$
can be summarized as follows. One observable is the total cross section, which
is related to it via the optical theorem
\be 
\sigma_{\rm tot}=4\pi \int db b \,{\rm  Im}\, \left( \Omega(s,b) \right) 
\ee
Some data points for the cross section are shown in 
the upper plot of Fig.~\ref{fig_pheno_from_profile6}. As  well known, the cross section
more than double from ISR to LHC energies. 
However, one should be aware that the low energy plots are affected by  the Reggeons other than the Pomeron.
Their contribution (from the PDG) is shown by the  the dotted line in the left corner.

Another important parameter is the elastic slope $B$ 
\be
B(t=0)=\left(-\frac{d{\rm ln}\sigma_e}{d|t|}\right)_0=\frac 12 \left<b^2\right>_\Omega
\ee
The corresponding data are shown  in 
the middle plot of Fig.~\ref{fig_pheno_from_profile6}. Their value
also grows with the collision energy, which is due to the effective growth
of the proton size induced by Gribov diffusion process. 

In order to understand what these data tell us about the shape of the
profile, it is useful also to plot their dimensionless ratio 

\be 
\label{RATIO}
R\equiv {\sigma_{\rm tot}(s) \over B(s)} 
\ee
with the cross section expressed in ${\rm GeV}^{-2}$. This ratio is plotted in the lower plot in 
Fig.~\ref{fig_pheno_from_profile6}. 
In order to understand what the ratio (\ref{RATIO}) tells us, it is convenient to compare it to some simplified
models for the shape,  such as e.g. the {\em black disc}

\be \Omega_{bd}=\theta(b_{max}-b) , \, \,\,R_{bd}=8\pi \ee
the $Gaussian$ without a prefactor
\be \Omega_{G}=e^{-b^2/2b_0^2}, \, \,\,R_{G}=4\pi \ee
and the $exponential$, also without a prefactor
\be  \Omega_{E}=e^{-b/b_0}, \, \,\,R_{E}={4\pi \over 3} \ee

The high energy values of the ratio are far from the two extreme shapes, and rather close
to the Gaussian value. The low energy profiles are in fact also near-Gaussian,
but with the prefactor smaller than one, as we will discuss below.

\subsection{The Pomeron shape and the mass of the string's ends}
As we already noted above, the 
shapes of the BFKL and BKYZ Pomeron scattering amplitudes $A(s,b)$ are both Gaussian, as  is
typical for diffusive processes.   The prefactor, however,  growing as $s^\Delta$,
at high enough  energy violates the unitarity.
% 
% is related to the observed differential elastic cross section
%through
%\be
%\label{VB50}
%\frac{d\sigma_e}{dt}=\frac 1{16\pi s^2} \left|\int d^2b\,e^{iq\cdot b}\,{\cal A}(\beta, b)\right|^2
%\ee 
%with ${\cal A}$ determined by one string exchange. In this approximation the  Pomeron
%%and similarly the BFKL Pomeron in the ladder approximation
%predicts  a  diffusive behavior (\ref{eqn_diff}) at large $b$. 
%In contrast, the data, both for the total cross section and show faster growth with energy, %consistent with a mean square size $\left<b^2\right>$ growing more rapidly than $ \chi$, in fact a growth 
%consistent with a combination of $\chi$ and $\chi^2$.

A generic resolution for this situation is well known: one has  
  to include the ``multi-Pomeron effect", see e.g.~\cite{Schegelsky:2011aa} and references therein. 
%  For a positive
% Pomeron intercept $\Delta_{\mathbb P}>0$,  the amplitude grows with energy 
%as $s^{\Delta_{\mathbb P}}$, and  when it reaches or exceeds one, one needs to ``unitarize" it.  
 A standard procedure  is  Glauber unitarization, which
follows from the substitution
\be
\label{GLAUBER}
 {\bf K} \rightarrow \Omega\equiv 1-e^{\bf -K}
\ee
%where the effective string propagator in impact parameter space is
%
%
%\be
%\label{001}
%{\bf K}=\frac{g^2_sa_D^2}{4\alpha^\prime}\left(\frac \pi{\chi}\right)^{\frac {D_\perp}2}e^{-{\cal S}(\beta, b)}
%\ee
%with ${\cal S}(\beta, b)$ as defined in (\ref{SPOT}). In the Appendix, we discuss the unitarization
%for the amplitudes of fixed charge conjugation signatures. 
%(\ref{001}) becomes of order one when ${\cal S}(\beta,b)$
%flips sign. This occure 
The two leading terms in ${\bf K}$ compensate each other when % ($\mu$ is small)
\be
 \chi \Delta \approx {b_{\rm black}^2 \over 4\chi \alpha^\prime_{\mathbb P}} 
\ee
For  $b< b_{\rm black}$ the interaction is too strong, 
multiple Pomeron exchanges  screen each other, and the proton is effectively black. 
%This effects essentially prevents us from 
%observing (and thus understanding) the amplitude at small $b$. The only way
%out is, as already mentioned above, to switch from proton to virtual photon collider,
%by dialing a very small dipole size $a_D$.
Asymptotically, we have $b_{\rm black}\sim \chi^2$.

%%%%%%%%%%%%%%%%%%%%

\begin{figure}
\begin{center}
\includegraphics[width=7cm]{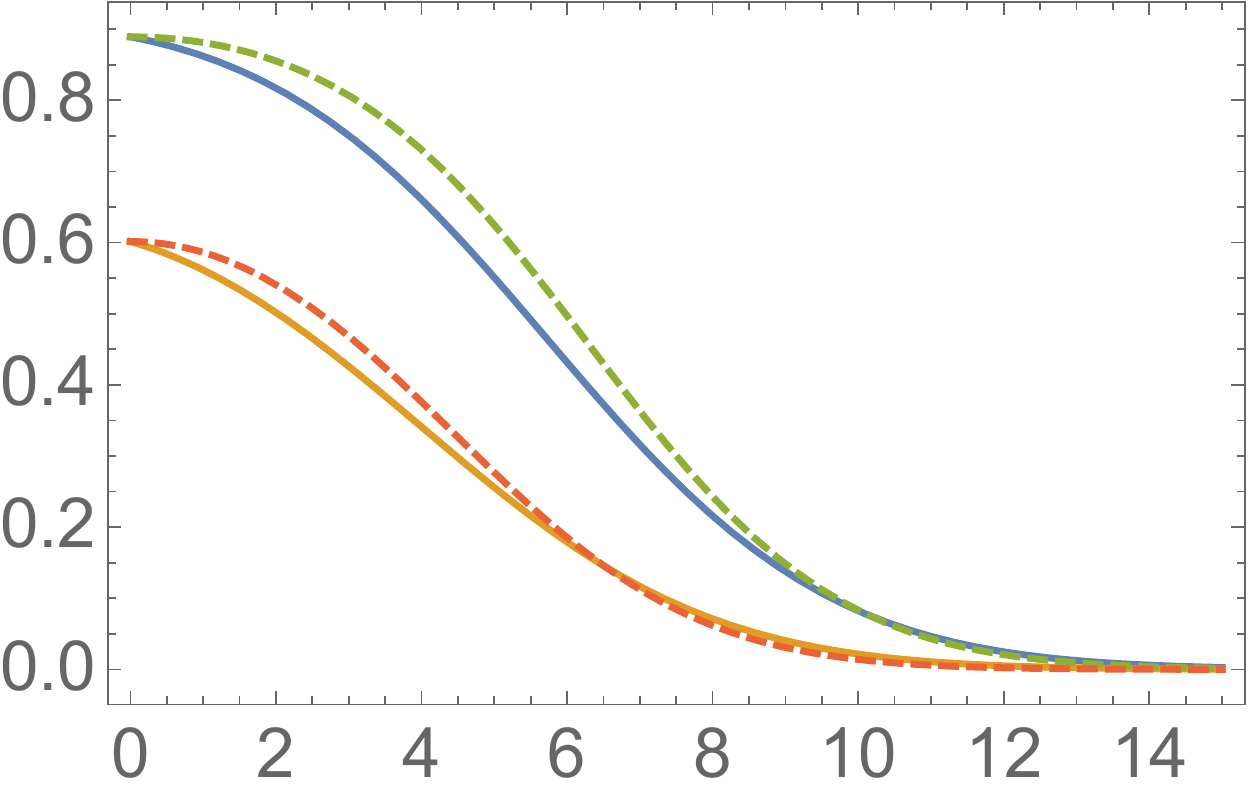}
\caption{The  unitarized profile function $\Omega=1-e^{\bf -K}$, for $\sqrt{s}=8000 \, {\rm GeV}$ (upper blue curve)
and $62\, {\rm GeV}$ (lower red curve), versus the impact parameter $b$ (${\rm GeV}^{-1}$).
The solid curves at all plots are for the parameters $A=7.\,{\rm GeV}^{-2}$, $\mu =-0.04\, {\rm GeV}$, and the dashed ones are for 
 $A=4.5\,{\rm GeV}^{-2}$, $\mu =0$. 
 The dotted line in the lower plot indicate the contribution of the Reggeons other than the Pomeron.
}
\label{fig_profile_from_profile6}
\end{center}
\end{figure}

The purpose of this subsection is to incorporate these effects, and also see
whether the fit to the data may indicate a nonzero value of the 
first higher order term $\mu$. A technical point is that 
standard Reggeon expressions for the amplitude contains
the nucleon form factors $f_N^2(t)$ times the Regge factor $s^{\alpha(t)}$.
However, in order to perform unitarization (accounting for multi-Pomerons) 
one has to proceed to the coordinate profile, and in this case the product involves the 
convolution of functions,  which is unnecessary complicated.
A simple way out, sufficient for the purpose of this section, is to use
Gaussian form factors  $f_N^2(q)=e^{-A q^2}$ 
which allows us to derive  the coordinate profile analytically

\bea
\label{KSB}
 {\bf K}(s,b)= C\,{\rm exp}\left(\Delta_{\mathbb P} \chi-{b^2 \over 4\left(A+\alpha^\prime_{\mathbb P} \chi\right)}+2\mu b\right)\nonumber\\
\eea
where the normalization constant $C=2.2$ is fitted to the TOTEM data, as all 
 higher order stringy contributions decreasing with $b$ are for now neglected.  
 We remark that while the occurence of $A$ in (\ref{KSB}) is exact for one-Pomeron
 exchange, it is only an approximation in the unitarized multi-Pomeron resummation
 in the Glauber substitution (\ref{GLAUBER}).

A negative $\mu$ shifts the maximum of the profile function away from $b=0$, and modifies the shape in a way shown 
(for the unitarized profile functions)  in Fig.~\ref{fig_profile_from_profile6}.
It turned out that the shape of the amplitude is sensitive to even small nonzero $\mu$. 

Now we invite the reader to look at Fig.~\ref{fig_pheno_from_profile6} again, now paying attention to the curves. In order to fit the upper plot, for the cross section, we used a smaller
Pomeron intercept $\Delta=0.09$. The middle plot, for the elastic slope, is however independent
of its value. 
%
%
% The dashed curve, with 
%Two other plots in this figure show the effect of the modification on the
% elastic slope parameter $B(\chi)$ and the inelastic cross section $\sigma_{in}(\chi)$
% as a function of $\chi=log(s/s_0)$. We ignored the elastic cross section since we
% have not included the real part in the amplitude in the simplified model under consideration.
% 
 One can see that the profile shape corresponding to $\mu=0$, the dashed line in the middle plot, describes the growth
 of  $B(\chi)$ quite well (but is not 
good for the value of the ratio at the highest LHC point $\sqrt{s}=7\, TeV$ shown.  
The solid line is better on that. The fact that it misses the low energy points
is less serious since one needs to subtract the non-Pomeron contributions.
 The upcoming LHC data for twice larger beam energy will clarify the situation, but for now
 we see that a small negative $\mu$ has a potential to make the description better.
   Perhaps the optimum,
 inside the simple ansatz for the shape we use, is somewhere in between these two curves. 
 
  The purpose of these comparisons is not to get an excellent fit to the data, as there are
  already several of those on the market, using multi-parameter and rather arbitrary
  functions. Our main message here is that the profile shape -- and the two observables reflecting   
 it --  are   very sensitive even to rather small $\mu$.  We
think this comparison tells us that its value is definitely small,  perhaps slightly negative.

There are also some theoretical justifications for a nonzero $\mu$, with a suggested sign.    In fact,  we already mentioned it in  (\ref{VB2}):
 the  extrinsic curvature term in the action does indeed produce such a contribution, see
~\cite{Hidaka:2009xh,Qian:2014jna}. 
One may further ask if lattice studies of the potential can also help to find 
the magnitude of $\mu$.
Unfortunately, on the lattice 
 the constant  term in the potential $\mu\sim 1/a$ is singular in the $a\rightarrow 0$ continuum limit
 (a pointlike charge). A subtraction is possible, but the accuracy of the remaining finite part is
 not  good enough.

   Concluding this section, let us also mention another
   (admittedly much more exotic) origin for the   $B(s),\sigma(s)$ growth with energy:
    a hypothetical $repulsive$ string self-interaction, briefly discussed in \cite{Shuryak:2017phz}.

%%%%%%%%%%%%%%%%%%%%%%%%%%%%%%
%%%%%%%%%%%%%%%%%%%%%%%%
%Before we start, one perhaps needs a bit of explanation of what is meant by the 
%size of the Pomeron. Of course, the beams actually used at colliders are 
%$p$ (or $\bar p$), and the proton has its own size (and form factors). 
%This is a complication which we prefer to ignore for now, considering
%a collision of some color dipoles of tunable size. In deep inelastic collisions
%(e.g. at future electron-ion collider EIC) one of the dipoles has size $1/Q$
%and thus it is small at large $Q$ and under experimental control. It would be
%even better to have a linear $e^+e^-$ collider in $\gamma^*\gamma^*$ mode,
%in which both dipoles are controllable: but for the moment we do not have those,
%and need to live with $pp$ collisions, with all their complications, such as the necessity to discuss the stringy Pomerons.
%
%

   \subsection{The  Pomeron profile and the Hagedorn transition}

In our paper~\cite{Shuryak:2013sra} we discussed the transition 
between the stringy and perturbative regime, using the so called Hagedorn transition
in which strings get highly excited. 
At the beginning of this paper we 
have discussed the static potential, in which a specific structure  -- the ``wiggle" indicating a transition between
two regimes -- has been identified. 
Furthermore, we argued that a stringy duality can connect the static potential
and the scattering amplitude, in a well defined way. 

Now, let us discuss the issue phenomenologically. Suppose a ``wiggle"
is also present in the Pomeron profile function: what observables should be
used to locate it? 

Our first comment is that $pp$ collisions at LHC energies already have a certain structure
in its unitarized profile function, namely the black disc. Obviously any
structure in the amplitude inside the black disc, for $r<r_{\rm bd}$, is unobservable.
To go around this difficulty one can either return to much lower energies of $pp$ collisions, 
or focus on $\gamma p$ collisions at the future Electron-Ion Collider.
 
Suppose now that the ``wiggle"  happens at impact parameters $outside$ the black disc
region. For simplicity, let us imagine a small peak in the profile function at certain 
value of the impact parameter $b_{\rm peak} > r_{\rm bd}$. A
Fourier-Bessel transform to momentum space will produce an oscillating signal
$\sim J_0(q b_{\rm peak})$. Its minimum is at $q b_{\rm peak}\approx 4$, and the
second peak is at $q b_{\rm peak}\approx 7$, not far from $2\pi$.

In fact TOTEM at 8 TeV does observe certain non-Gaussian deviations of the elastic peak, 
parametrized by a simple Gaussian $e^{{\rm const}* t}$, with  a minimum located
at $q^2\approx 0.1 \, {\rm GeV}^2$. If this corresponds to a structure in the profile function, 
its location should be at 
\be b_{\rm peak}={4 \over 0.316 \, {\rm GeV}}\approx 2.5 \, {\rm fm}\ee
This distance is  way too far for the location of the Hagedorn transition.
The corresponding ``tube" temperature is way too low, and cannot correpond
to this transition.

  \section{Summary}

The effective string theory suggests corrections to the fundamental string Lagrangian, and elucidates how these corrections modify the linearly rising potetial at shorter distances. QCD lattice numerical simulations do  indeed
reveal   two universal ``Luscher terms" %at shorter distances,
  and a new non-universal term at even shorter distances
responsible for the appearance   of a  ``wiggle" in  the potential. Albeit  small in amplitude,  it
clearly marks 
 the transition from the perturbative to the stringy regime.

The description of the static heavy quark potential in QCD can be mapped onto 
the Pomeron scattering amplitude by string duality. 
We have suggested that  the scattering amplitude can provide  a more direct
 test of the string dynamics than the static potential. We have discussed the effects predicted
 by those approaches, for the Pomeron size and shape.  The future experiments at the Electron-Ion Collider should allow,
 through the measurements of differential cross sections of $\gamma^* p$ scattering,
  to elucidate the structure of the Pomeron and 
 more accurately define the domains of its pQCD and string descriptions.

  \vskip 1cm
{\bf Acknowledgements.} We thank V.A.Petrov for useful comments on the first version of this manuscript. 
This work was supported by the U.S. Department of Energy under Contracts No. DE-FG-88ER40388 and DE-AC02-98CH10886.
\vskip 0.5cm

\section{Appendix: Unitarization for fixed signature}

The fully unitarized scattering amplitudes of given charge conjugation signature ${\cal T}^{C=\pm}$
in impact parameter space, follows from (\ref{VB3}) in terms of ${\bf K}$,  and its charge conjugate $\chi\rightarrow \chi-i\pi$
in terms of $\tilde{\bf K}$.  Specifically, for fixed impact parameter we have

\bea
\label{APP1}
&&{\cal T}^++{\cal T}^-=2is\left(1-e^{-{\bf K}+\tilde{\bf K}}\right)\nonumber\\
&&{\cal T}^+-{\cal T}^-=2is\left(1-e^{-{\bf K}-\tilde{\bf K}}\right)
\eea
 or equivalently

\bea
&&{\cal T}^+=+2is\left(1-e^{-{\bf K}}{\rm cosh}\,\tilde{\bf K}\right)\nonumber\\
&&{\cal T}^-=-2is\,e^{-{\bf K}}\,{\rm sinh}\,\tilde{\bf K}
\eea
The total cross sections for $pp$ and $p\bar p$ scattering can be deduced  
from the forward scattering parts of (\ref{APP1}) 

\bea
\label{APP2}
&&\frac 12 \left(\sigma_{p\bar p}+\sigma_{pp}\right)=\frac {4\pi}s \,{\rm Im}\,\tilde{\cal T}^+\nonumber\\
&&\frac 12 \left(\sigma_{p\bar p}-\sigma_{pp}\right)=\frac {4\pi}s \,{\rm Im}\,\tilde{\cal T}^-
\eea
where the Fourier transform at $t=0$  is subsumed in the amplitudes in (\ref{APP2}).


\begin{thebibliography}{99} 

 
%\cite{Gribov:1973jg}
\bibitem{Gribov:1973jg} 
  V.~N.~Gribov,
  %``Space-time description of hadron interactions at high-energies,''
  hep-ph/0006158;
  %%CITATION = HEP-PH/0006158;%%
  %73 citations counted in INSPIRE as of 15 Jun 2017
V.~N.~Gribov, 'Gauge Theories and Quark Confinement', 2002, PHASIS.



%\cite{Kuraev:1977fs}
\bibitem{Kuraev:1977fs} 
  E.~A.~Kuraev, L.~N.~Lipatov and V.~S.~Fadin,
  %``The Pomeranchuk Singularity in Nonabelian Gauge Theories,''
  Sov.\ Phys.\ JETP {\bf 45}, 199 (1977)
  [Zh.\ Eksp.\ Teor.\ Fiz.\  {\bf 72}, 377 (1977)];
  %%CITATION=SPHJA,45,199;%%
  %2922 citations counted in INSPIRE as of 13 Jun 2017
%\cite{Balitsky:1978ic}
%\bibitem{Balitsky:1978ic} 
  I.~I.~Balitsky and L.~N.~Lipatov,
  %``The Pomeranchuk Singularity in Quantum Chromodynamics,''
  Sov.\ J.\ Nucl.\ Phys.\  {\bf 28}, 822 (1978)
  [Yad.\ Fiz.\  {\bf 28}, 1597 (1978)].
  %%CITATION=SJNCA,28,822;%%
  %3224 citations counted in INSPIRE as of 13 Jun 2017


\bibitem{PIRNER}
%\cite{Shoshi:2002in}
%\bibitem{Shoshi:2002in} 
  A.~I.~Shoshi, F.~D.~Steffen and H.~J.~Pirner,
  %``S matrix unitarity, impact parameter profiles, gluon saturation and high-energy scattering,''
  Nucl.\ Phys.\ A {\bf 709}, 131 (2002)
 % doi:10.1016/S0375-9474(02)01042-4
  [hep-ph/0202012].
  %%CITATION = doi:10.1016/S0375-9474(02)01042-4;%%
  %87 citations counted in INSPIRE as of 02 Sep 2017

%\cite{Brower:2006ea}
\bibitem{Brower:2006ea} 
  R.~C.~Brower, J.~Polchinski, M.~J.~Strassler and C.~-ITan,
  %``The Pomeron and gauge/string duality,''
  JHEP {\bf 0712}, 005 (2007)
  [hep-th/0603115].
  %%CITATION=HEP-TH/0603115;%%
%\cite{Brower:2007xg}
%\bibitem{Brower:2007xg} 
  R.~C.~Brower, M.~J.~Strassler and C.~-ITan,
  %``On The Pomeron at Large 't Hooft Coupling,''
  JHEP {\bf 0903}, 092 (2009)
  [arXiv:0710.4378 [hep-th]].
  %%CITATION=ARXIV:0710.4378;%%
  



  %\cite{Basar:2012jb}
\bibitem{Basar:2012jb} 
  G.~Basar, D.~E.~Kharzeev, H.~U.~Yee and I.~Zahed,
  %``Holographic Pomeron and the Schwinger Mechanism,''
  Phys.\ Rev.\ D {\bf 85}, 105005 (2012)
%  doi:10.1103/PhysRevD.85.105005
  [arXiv:1202.0831 [hep-th]].
  %%CITATION = doi:10.1103/PhysRevD.85.105005;%%
  %25 citations counted in INSPIRE as of 12 Jul 2017


%\cite{Shuryak:2013sra}
\bibitem{Shuryak:2013sra} 
  E.~Shuryak and I.~Zahed,
  %``New regimes of the stringy (holographic) Pomeron and high-multiplicity $pp$ and $pA$ collisions,''
  Phys.\ Rev.\ D {\bf 89}, no. 9, 094001 (2014)
 % doi:10.1103/PhysRevD.89.094001
  [arXiv:1311.0836 [hep-ph]].
  %%CITATION=doi:10.1103/PhysRevD.89.094001;%%
  %31 citations counted in INSPIRE as of 13 Jun 2017

%\cite{Polyakov:1986cs}
\bibitem{Polyakov:1986cs} 
  A.~M.~Polyakov,
  %``Fine Structure of Strings,''
  Nucl.\ Phys.\ B {\bf 268}, 406 (1986);
  doi:10.1016/0550-3213(86)90162-8
  %%CITATION = doi:10.1016/0550-3213(86)90162-8;%%
  %577 citations counted in INSPIRE as of 02 Aug 2017

\bibitem{Luscher}
M. Luscher, Nucl. Phys. B 180, 317 (1981); for a review see J. Greensite, Prog. Part. Nucl. Phys. 51, 1 (2003) [arXiv:hep-lat/0301023].




%\cite{Luscher:2004ib}
\bibitem{Luscher:2004ib} 
  M.~Luscher and P.~Weisz,
  %``String excitation energies in SU(N) gauge theories beyond the free-string approximation,''
  JHEP {\bf 0407}, 014 (2004)
 % doi:10.1088/1126-6708/2004/07/014
  [hep-th/0406205].
  %%CITATION = doi:10.1088/1126-6708/2004/07/014;%%
  %144 citations counted in INSPIRE as of 11 Jul 2017


%\cite{Arvis:1983fp}
\bibitem{Arvis:1983fp} 
  J.~F.~Arvis,
  %``The Exact $q \bar{q}$ Potential in Nambu String Theory,''
  Phys.\ Lett.\  {\bf 127B}, 106 (1983).
 % doi:10.1016/0370-2693(83)91640-4
  %%CITATION = doi:10.1016/0370-2693(83)91640-4;%%
  %184 citations counted in INSPIRE as of 15 Jun 2017

%\cite{Aharony:2010db}
\bibitem{Aharony:2010db} 
  O.~Aharony and N.~Klinghoffer,
  %``Corrections to Nambu-Goto energy levels from the effective string action,''
  JHEP {\bf 1012}, 058 (2010)
  %doi:10.1007/JHEP12(2010)058
  [arXiv:1008.2648 [hep-th]].
  %%CITATION = doi:10.1007/JHEP12(2010)058;%%
  %43 citations counted in INSPIRE as of 11 Jul 2017



 %\cite{Hidaka:2009xh}
\bibitem{Hidaka:2009xh} 
  Y.~Hidaka and R.~D.~Pisarski,
  %``Zero Point Energy of Renormalized Wilson Loops,''
  Phys.\ Rev.\ D {\bf 80}, 074504 (2009)
  %doi:10.1103/PhysRevD.80.074504
  [arXiv:0907.4609 [hep-ph]].
  %%CITATION = doi:10.1103/PhysRevD.80.074504;%%
  %21 citations counted in INSPIRE as of 11 Jul 2017
  
  
  
  %\cite{Qian:2014jna}
\bibitem{Qian:2014jna} 
  Y.~Qian and I.~Zahed,
  %``A Stringy (Holographic) Pomeron with Extrinsic Curvature,''
  Phys.\ Rev.\ D {\bf 92}, no. 8, 085012 (2015)
  %doi:10.1103/PhysRevD.92.085012
  [arXiv:1410.1092 [nucl-th]].
  %%CITATION = doi:10.1103/PhysRevD.92.085012;%%
  %3 citations counted in INSPIRE as of 11 Jul 2017
  

 \bibitem{footnote}
  In~\cite{Qian:2014jna} this shift effect was found to be suppressed as $\mu b/\chi$.
  The  expansion used upsets the string  duality enforced here. 




%\cite{Petrov:2014jya}
\bibitem{Petrov:2014jya} 
  V.~A.~Petrov and R.~A.~Ryutin,
  %``High-energy scattering versus static QCD strings,''
  Mod.\ Phys.\ Lett.\ A {\bf 30}, no. 18, 1550081 (2015)
  doi:10.1142/S0217732315500819
  [arXiv:1409.8425 [hep-ph]].
  %%CITATION = doi:10.1142/S0217732315500819;%%
  %1 citations counted in INSPIRE as of 10 Oct 2017


  
  %\cite{Brandt:2017yzw}
\bibitem{Brandt:2017yzw} 
  B.~B.~Brandt,
  %``Spectrum of the open QCD flux tube and its effective string description I: 3d static potential in SU(N = 2, 3),''
  JHEP {\bf 1707}, 008 (2017)
  doi:10.1007/JHEP07(2017)008
  [arXiv:1705.03828 [hep-lat]].
  %%CITATION = doi:10.1007/JHEP07(2017)008;%%

 




%\cite{Shuryak:2017phz}
\bibitem{Shuryak:2017phz} 
  E.~Shuryak and I.~Zahed,
  %``Regimes of the Pomeron and its Intrinsic Entropy,''
  arXiv:1707.01885 [hep-ph].
  %%CITATION = ARXIV:1707.01885;%%
  


  
 \bibitem{STOFFERS} 
  %\cite{Stoffers:2012zw}
%\bibitem{Stoffers:2012zw} 
  A.~Stoffers and I.~Zahed,
  %``Holographic Pomeron: Saturation and DIS,''
  Phys.\ Rev.\ D {\bf 87}, 075023 (2013)
  %doi:10.1103/PhysRevD.87.075023
  [arXiv:1205.3223 [hep-ph]].
  %%CITATION = doi:10.1103/PhysRevD.87.075023;%%
  %14 citations counted in INSPIRE as of 02 Sep 2017


%\cite{Burq:1982ja}
\bibitem{Burq:1982ja} 
  J.~P.~Burq {\it et al.},
  %``Soft $\pi^- p$ and $p p$ Elastic Scattering in the Energy Range 30-{GeV} to 345-{GeV},''
  Nucl.\ Phys.\ B {\bf 217}, 285 (1983).
 % doi:10.1016/0550-3213(83)90149-9
  %%CITATION = doi:10.1016/0550-3213(83)90149-9;%%
  %70 citations counted in INSPIRE as of 13 Jul 2017



  %\cite{Donnachie:1992ny}
\bibitem{Donnachie:1992ny} 
  A.~Donnachie and P.~V.~Landshoff,
  %``Total cross-sections,''
  Phys.\ Lett.\ B {\bf 296}, 227 (1992)
  [hep-ph/9209205].
  %%CITATION=HEP-PH/9209205;%%






 %\cite{Meyer:2004gx}
\bibitem{Meyer:2004gx} 
  H.~B.~Meyer,
  %``Glueball regge trajectories,''
  hep-lat/0508002.
  %%CITATION = HEP-LAT/0508002;%%
  %90 citations counted in INSPIRE as of 27 Jul 2017


%\cite{Antchev:2016vpy}
\bibitem{Antchev:2016vpy} 
  G.~Antchev {\it et al.} [TOTEM Collaboration],
  %``Measurement of elastic pp scattering at $\sqrt{\hbox {s}} = \hbox {8}$  TeV in the Coulombï¿œnuclear interference %region: determination of the $\mathbf {\rho }$ -parameter and the total cross-section,''
  Eur.\ Phys.\ J.\ C {\bf 76}, no. 12, 661 (2016)
%  doi:10.1140/epjc/s10052-016-4399-8
  [arXiv:1610.00603 [nucl-ex]];
  %%CITATION = doi:10.1140/epjc/s10052-016-4399-8;%%
  %23 citations counted in INSPIRE as of 28 Jul 2017
%\cite{Antchev:2013gaa}
%\bibitem{Antchev:2013gaa} 
  G.~Antchev {\it et al.} [TOTEM Collaboration],
  %``Measurement of proton-proton elastic scattering and total cross-section at S**(1/2) = 7-TeV,''
  Europhys.\ Lett.\  {\bf 101}, 21002 (2013).
  %doi:10.1209/0295-5075/101/21002
  %%CITATION = doi:10.1209/0295-5075/101/21002;%%
  %156 citations counted in INSPIRE as of 13 Jul 2017



%\cite{Schegelsky:2011aa}
\bibitem{Schegelsky:2011aa} 
  V.~A.~Schegelsky and M.~G.~Ryskin,
  %``The diffraction cone shrinkage speed up with the collision energy,''
  Phys.\ Rev.\ D {\bf 85}, 094024 (2012)
%  doi:10.1103/PhysRevD.85.094024
  [arXiv:1112.3243 [hep-ph]].
  %%CITATION = doi:10.1103/PhysRevD.85.094024;%%
  %30 citations counted in INSPIRE as of 13 Jul 2017



  












  %\cite{Bakry:2017utr}
%\bibitem{Bakry:2017utr} 
 % A.~Bakry, X.~Chen, M.~Deliyergiyev, A.~Galal, S.~Xu and P.~M.~Zhang,
  %``Stiff self-interacting string near QCD deconfinement point,''
 % arXiv:1707.02962 [hep-lat].
  %%CITATION = ARXIV:1707.02962;%%









\end{thebibliography}
\end{document}